\newcommand{\beq}{\begin{equation}}
\newcommand{\enq}{\end{equation}}
\newcommand{\beqarr}{\begin{eqnarray}}
\newcommand{\enqarr}{\end{eqnarray}}
\newcommand{\eqref}[1]{Eq.\ (\ref{#1})}
\newcommand{\figref}[1]{Fig.\ (\ref{#1})}
\newcommand{\ABS}[1]{\left| #1 \right|}
\newcommand{\AVG}[1]{\left< #1 \right>}
\newcommand{\lesim}{\stackrel{<}{\scriptstyle \sim}}
\newcommand{\PRL}[1]{Phys.\ Rev.\ Lett. {\bf #1}}
\newcommand{\PRB}[1]{Phys.\ Rev.\ B {\bf #1}}
\newcommand{\PRA}[1]{Phys.\ Rev.\ A {\bf #1}}

\newcommand{\Et}{E_{T}}
\newcommand{\jcdw}{j_{\rm CDW}}
\newcommand{\kb}{k_{B}}
\newcommand{\phii}{\phi_{i}}
\newcommand{\phij}{\phi_{j}}
\newcommand{\beti}{\beta_{i}}
\newcommand{\eti}{\eta_{i}}
\newcommand{\sik}{\psi_{k}}
\newcommand{\bk}{B_{k}}
\newcommand{\chik}{\chi_{k}}
\newcommand{\bbar}{\overline{B}}
\newcommand{\ra}{\rightarrow}

\documentstyle[preprint,revtex]{aps}
\begin{document}
\begin{title}
Thermal Rounding of the Charge Density Wave\\
Depinning Transition
\end{title}

\author{A. Alan Middleton}
\begin{instit}
Physics Department, Syracuse University, Syracuse, NY 13244
\end{instit}
\receipt{}

\begin{abstract}
The rounding of the charge density wave depinning transition
by thermal noise is examined.
Hops by localized modes over small barriers
trigger ``avalanches'', resulting in a creep velocity
much larger than that expected from comparing
thermal energies with typical barriers.
For a field equal to the $T=0$ depinning field, the creep
velocity is predicted to have a {\em power-law} dependence on the
temperature $T$;
numerical computations confirm this result.
The predicted order of magnitude of the thermal rounding of the
depinning transition is consistent with rounding seen in experiment.
\end{abstract}

The model \cite{flrscf} of an incommensurate charge density wave (CDW) as
a deformable medium has been successful in explaining many
experimental results both qualitatively and quantitatively \cite{review}.
In this model, the CDW is an elastic medium
subject both to impurity pinning forces and an external drive force.
At zero temperature, there is a sharp depinning transition:
when the applied electric field $E$
exceeds a threshold value $\Et$, the CDW depins, and the sliding CDW
carries an electric current $\jcdw$.
Fisher \cite{dsfone} has proposed that the behavior near the
depinning transition can be understood as a type of dynamic critical
phenomenon.
In particular, the CDW current $\jcdw$ should behave as
$\jcdw \sim (E-\Et)^\zeta$, with $\zeta$ a critical exponent.
Numerical work \cite{numerics} has
confirmed aspects of this picture in finite-dimensional models, and values
for the exponent $\zeta$ have been determined \cite{thesis,myersethna}.

This same numerical work has shown that, in order to clearly see the dynamic
critical behavior, it is necessary to examine fields within a few percent
of the depinning field $\Et$.
It has proven difficult
to obtain clear experimental results for applied fields this near to the
depinning transition.
A major cause of this difficulty is that the transition between the
pinned and sliding state is not perfectly sharp
\cite{bhatt} ; there is an apparent
rounding in the $\jcdw(E)$ relationship.
Besides affecting the measurement of currents for $E$ near $\Et$, the
rounding makes it difficult to determine the threshold
field $\Et$ accurately.
In order to make a careful comparison between experiment and
theory, it is necessary to understand the rounding of the depinning
transition.

Various explanations for this rounding have been proposed.
One possibility is that experimental samples are
macroscopically inhomogeneous, with
local regions depinning at distinct applied fields $E$.
The bulk $\jcdw(E)$ relation would then be smeared.
However, the sharp narrow band noise
seen in many samples \cite{review},
indicates that the whole CDW is sliding at a uniform rate
(though the narrow band noise is a finite size effect, its
presence in finite experimental samples indicates uniform velocity).
A somewhat related explanation is that provided by phase slip
\cite{pslip} in macroscopically homogeneous samples.
Coppersmith has shown \cite{sue} that there exist
exponentially rare regions with atypically weak pinning.
Phase slip occurring on the border of such regions allows for an
excess current below the bulk depinning field $\Et$.
The main difficulty in attributing the rounding of the transition
to this effect is the rarity of these atypical regions \cite{nbnnote}.
In particular, the magnitude of this effect will decrease as the
exponential of a negative power of the impurity concentration.
It has also been proposed \cite{bhatt} that thermal
noise may round the CDW depinning transition.

In this paper, I examine the effects of finite temperature $T$ on
the CDW depinning transition.
Recent work \cite{natterman} has
examined thermal effects at small field, far from the transition, where
a creep velocity $\propto \exp[(ET)^{-\mu}]$ is expected, for some
exponent $\mu$.
I study here instead the thermal effects at fields $E \approx \Et$,
by examining the barriers that prevent the forward motion of
Lee-Rice domains (regions of a size such that pinning and elastic
forces are comparable and that act roughly as single degrees
of freedom \cite{flrscf}).
At first sight, thermal effects in CDW's might be expected to be very
small, since the
thermal energies are $\sim 10^{3}-10^{7}$ times smaller
than the typical barrier energies
estimated from the magnitude of the threshold
field.
This is consistent with the thermal creep for small fields, $E \ll \Et$, being
extremely small.
However, at $T=0$, for fields just below
the depinning threshold, many of the barriers to forward
motion are much smaller than the typical barriers \cite{dsfone,pbl}.
If the field is increased slightly, the smallest barriers
vanish, and spatially localized instabilities are
induced, leading to the ``jumping'' forward of individual Lee-Rice
domains.
Numerical calculations at $T=0$
for fields below threshold show that the jumping forward of a single domain
may cause an ``avalanche'', in which a large region of the CDW slides
forward, as shown in Refs.\ \cite{thesis,myersethna}.
At fields just above threshold, there is no stable
configuration, and these triggering jumps and resulting
avalanches lead to a positive CDW velocity.
The argument I develop here estimates the current induced by the
inclusion of thermal noise, which causes ``hops'' with effects similar to
those due to the ``jumps'' resulting from an increase in field.
These ``hops'' result in avalanches, as shown in Fig.~1.
For $E=\Et$, this argument results in a prediction for the CDW velocity
(directly proportional to the current $\jcdw$) of the form
\beq
\label{eq:scaleone}
v(T,E=\Et) \sim T^{\zeta/\tau},
\enq
where $\tau$ is a non-universal exponent, with the values $\tau=3/2$ for
the usual Fukuyama-Lee-Rice model, and $\tau=2$ for a ``ratcheted-kick''
model \cite{thesis}.
The form of this scaling (with $\tau=3/2$) is the same as that predicted
by Fisher in mean field theory using a related, but distinct, argument
\cite{dsfone}.
I find the power law behavior (\ref{eq:scaleone}) to be consistent
with numerical results in dimensions $d=2,3$.
For thermal energies
on the order of $10^{-4}$ of the typical barrier height,
these numerical calculations show a broadening of the transition
on the order of 0.5\%, in $d=2,3$,
in order of magnitude agreement with experimental observations.

The Fukuyama-Lee-Rice model Hamiltonian \cite{flrscf}
on a cubic (or square) lattice in scaled units
is given by \cite{numerics,foot:units,pbl}:
\beq
\label{eq:energy}
{\cal H} = \frac{1}{2}\sum_{\left< i j \right>} (\phii-\phij)^{2}
+h \sum_{i} V(\phii-\beti)-E\sum_{i}\phii,
\enq
where $\phii$ is the CDW phase at spatial site $i$,
$h$ is the impurity pinning strength,
$\beti$ is a random pinning phase, uniformly chosen in the interval
$[0,2\pi)$, and $V(\phi-\beta)$ gives the shape of the pinning potential.
The first term of \eqref{eq:energy} represents elastic interactions between
nearest neighbor pairs of sites
$\left< i j \right>$, the second term models pinning
by impurities, and the last term represents the CDW polarization energy.
I consider here two potentials $V$: the usual, smooth potential
$V=\cos(\phii-\beti)$ and a sharp ``ratcheted-kick'' potential \cite{thesis},
which is, effectively, a sawtooth function of $\phii-\beti$.
Evolution of the $\phii$ in time $t$ is given by overdamped equations of
motion $d\phii/dt=-(\partial {\cal H}/\partial \phii)+\eti(t)$,
with Langevin noise $\eti(t)$,
$\left< \eti(t) \eti(t') \right>=4T\Et\delta(t-t')$, so that
temperature $T$ is measured in units of the barrier energy scale
$2\Et$ set by the threshold field.
For $T=0$, the velocity $v=\left<d\phii/dt\right>_{i}$ is zero for $\ABS{E}
\leq \Et$ and is positive for $E > \Et$,
with $v \sim f^\zeta$ for small reduced field
$f \equiv (E-\Et)/\Et$.
As there is no phase slip in this model, the temporally averaged
velocity must be uniform
throughout the system, and there is a {\em single} threshold field at $T=0$.
For $\ABS{E} < \Et$, the polarization is defined as $P \equiv
\AVG{\phii}$.
Numerical investigation of the sub-threshold state \cite{thesis,pbl,sncdsf}
shows that the {\em
linear} polarizability of the CDW can be understood as the sum of
contributions from spatially localized modes.
As the applied field is increased, the modes with the smallest linear
relaxation rates become unstable (via a saddle-node bifurcation),
resulting in localized, irreversible jumps
in the CDW phase.
These jumps trigger avalanches over regions which become larger as threshold
is approached; these jump and avalanche combinations are the
dominant contribution to the {\em nonlinear}
polarizability of the CDW \cite{thesis}.

Consider the CDW configuration $\{\phii\}$ at the threshold field, $E=\Et$,
and temperature $T=0$.
The localized modes of the CDW configuration can be considered as (linearly)
independent degrees  of freedom $\sik$ ($k$ being mode indices),
each with size approximately that of a Lee-Rice domain.
Each $\sik$ sees an individual
potential $W_{k}(\sik)$, due to elastic, pinning, and drive forces, with
$\sum W_{k}(\sik)={\cal H}$.
When the CDW is in a stationary configuration,
each $\sik$ is at a local minimum of $W_{k}$.
Expanding $W_{k}$ in powers of $\sik$,
with $\sik \equiv 0$ at the inflection point
separating the local minimum from the barrier to forward motion, gives
(see also Refs.\ \cite{dsfone,numerics}):
\beq
\label{eq:pot}
W(\sik)=\frac{1}{3}A\sik^{3}+\bk\sik-(\chik+1)(E-\Et)\sik+O(\sik^{5}),
\enq
where $A$ is approximately independent of $k$ with a value determined by
the strength and periodicity of the pinning potential and $\bk$ and $\chik$
are positive;
for the softest modes $k$, $\bk$ approaches zero, while $\chik$, which
incorporates the linear polarizability of neighboring domains, is of order
one.
The potential $W(\sik)$ changes in
response to an increase of the applied field.
If $E$ is increased by a small amount $\delta E$, minima of the local
potentials vanish where $\bk$ is smaller than $\delta E/(\chik+1)$.
As a result, some $\sik$ jump forward and trigger an avalanche.
By {\em destabilizing} a local mode, the increase of the field $E$
leads to a current pulse.
I now argue that, using \eqref{eq:pot},
the effects of hops caused by a finite temperature
can be directly related to the effects of the field induced
destabilization of local modes.

It is useful to note two numerical results from $T=0$ simulations
of CDW's in finite dimensions:
the apparent insensitivity
of the exponent $\zeta$ to the shape of the pinning
potential and the relatively quick relaxation of the velocity in the moving
state.
In the rather different cases of the smooth pinning potential and
the ratcheted-kick potential, $\zeta$ is found to be universal to within
numerical error, for $d=2$, even though the mean-field exponents are
distinct.
The difference in the mean-field critical behavior
of the two models results from the details of how a jump occurs.
The apparent universality of $\zeta$ therefore
suggests that the details of {\em how}
a single domain jumps is unimportant
in determining the critical behavior of the velocity
in finite-dimensional models.
This conclusion is supported by the observation that
the velocity equilibrates well before the configuration reaches
a periodic state.
These numerical observations imply that the
velocity is dominated by the
avalanches which are triggered by the jumps, with the critical
behavior of the velocity independent of jump dynamics and the
shape of the sliding configuration.

These results suggest that thermally-induced
hopping over small barriers will
have the same effect on the velocity as raising the applied
field by an amount which causes the destabilization of the same soft modes,
if the time scale for the hop is much smaller than that for the evolution
of the resulting avalanche.
For potentials of the form \eqref{eq:pot}, the barrier height $\Delta_{k}$
behaves as $\Delta_{k} \sim \bk^{3/2}$.
Since a finite temperature $T$ will
cause rapid tunneling over barriers of height $\Delta_{k} \approx T$,
thermal noise triggers avalanches at a rate corresponding
to an applied field $E$ with $(E-\Et) \sim T^{2/3}$,
assuming $\chik$ of order one \cite{nanote}.
It follows that, for small $T$ at an applied field
$E=\Et$, the velocity $v$, due to avalanches triggered by
thermal noise, scales as $v\sim  T^{\zeta/\tau},$
where the exponent $\tau=3/2$.
For the case where
pinning potential $V(\phi-\beta)$ is given by the
ratcheted-kick form, it can be similarly shown that $\tau=2$.
Thermal effects are therefore non-universal in the shape of the pinning
potential
(it is probable however that the thermal effects {\em are}
universal if the noise is due to finite
random kicks, rather than Langevin noise \cite{thesis}).
The comparison of the effects of small reduced fields $f$ and temperature
$T$ also suggests the scaling form
\beq
\label{eq:scaletwo}
v(f,T)=T^{\zeta/\tau}\bbar(fT^{-1/\tau}),
\enq
which is exactly the form proposed by Fisher for
mean-field theory \cite{dsfone},
with $\bbar(u)$ behaving as $u^{\zeta}$ for large positive $u$ and
decaying very rapidly as $u\ra -\infty$.

I have used
numerical simulations for dimensions $d=2,3$ to check the
scaling behavior predicted by
Eqs.\ (\ref{eq:scaleone},\ref{eq:scaletwo}).
The pinning strength $h=(2.5)d$ is
chosen so that the size of a Lee-Rice domain is
approximately one lattice unit.
The computations were performed on a Connection Machine CM-2, utilizing
16K processors for approximately 35h.

Fig.~2 shows velocity $v$ as a function of dimensionless
temperature $T$ on a log-log scale,
for fixed field $E=\Et$ and a smooth potential $V$.
The straight lines shown are not fits to the data, but have slopes
predicted by \eqref{eq:scaleone} with values for
the dynamical exponent of \cite{thesis,myersethna}
$\zeta=0.63, 0.85$ in
$d=2,3$, respectively (estimated error bars for $\zeta$ in $d=2,3$ are
0.07; the value for $\zeta$ in $d=3$ is from simulations on the
ratcheted-kick model), and the predicted exponent $\tau=3/2$.
There is very good agreement between the data and \eqref{eq:scaleone},
for velocities $v \lesim 0.1$ (this velocity scale corresponds to the
crossover between the high field and critical behavior in $T=0$ simulations
\cite{thesis}.)
If the slope is allowed to vary, a best fit yields $\zeta/\tau=0.40\pm0.04,
0.60\pm0.06$ in $d=2,3$, respectively.
Similar calculations for the ``ratcheted-kick'' model in $d=2,3$
also show power-law behavior consistent with (\ref{eq:scaleone}),
for the above values of $\zeta$, but with $\tau=2$.
It  follows that the relationship between $v$ and $\kb T/h$ is
non-universal; this is not surprising given the derivation of
\eqref{eq:scaleone}, which, when Langevin noise is used,
depends on microscopic properties of the model.

In order to examine the (non-universal) scaling function $\bbar$ of
\eqref{eq:scaletwo} for a particular case, velocities have also
been calculated for fixed temperature and varying field.
In \figref{fig:scalingfn}, scaled velocity $vT^{-\zeta/\tau}$ is plotted as a
function of scaled reduced field $fT^{-1/\tau}$ for the smooth pinning
potential in $d=2$, for the above values of $\zeta$ and $\tau$.
The scaling form \eqref{eq:scaletwo} describes the data
well.
If $\zeta$ and $\tau$ are allowed to vary so as to produce the best
fit to a single curve, I find $\zeta=0.6\pm0.1$ and $\tau=1.6\pm0.2$
(with subjective error bars).
For dimensionless temperatures $T$ of order $10^{-4}$,
this plot indicates that
the rounding of the transition occurs over a range of $\sim1\%$ in reduced
field, in $d=2$.
In general, the width in field $E$ of the rounding of the $v$ vs.\ $E$ curve,
will be some constant times $\Et T^{1/\tau}$, independent of dimension.
Using this criterion and the data of \figref{fig:smooth}, the
width of the rounding is found to be $\sim 0.5\%$ for $T=10^{-4}$ in $d=3$.

The scaling predictions of Eqs.\
(\ref{eq:scaleone},\ref{eq:scaletwo}), which are
supported by these numerical simulations,
can be directly applied to experiment.
The temperature scale set by the typical barrier is of the order
$E_{B}=2\rho V_{\rm FLR} \Et \lambda_{\rm CDW}$,
where $\lambda_{\rm CDW}$ is the wavelength of the CDW, which
determines the periodicity of the pinning potential, $\rho$ is the CDW
charge density, and $V_{\rm FLR}$ is the Fukuyama-Lee-Rice domain
volume.
For typical values of the parameters in ${\rm NbSe}_{3}$
($\lambda_{\rm CDW}=4\AA$, $\Et=100$ mV/cm, $\rho=2\times 10^{5} {\rm
C/m^{3}}$, $V_{\rm FLR}=(0.1\mu{\rm m})^{3}$), at 125K, the
dimensionless temperature is $T= 3 \times 10^{-5}$, on the order of the
smallest temperatures examined in the simulations reported here,
so that the rounding in this case should be on the order of 1\%.
Generally, the materials parameters, such as the elastic constant (which
affects $\xi_{\rm FLR}$) and charge density, are strongly dependent on
temperature, making it quite difficult to verify the scaling predictions
by comparing with experiments where the physical temperature is varied.
The most direct method for checking the behaviors in
Eqs.\ (\ref{eq:scaleone},\ref{eq:scaletwo})
may be to control the {\em dimensionless} temperature by varying the
concentration of weak pinning impurities.
For weak pinning, with impurity concentration $n_{i}$, the dimensionless
temperature $T\propto E_{B}^{-1} \propto n_{i}^{(2-d)/(d-4)}$
\cite{flrscf,review}.

I am pleased to acknowledge discussions with D.~S.~Fisher,
P.~B.~Littlewood, J.~Sethna, and C.~Myers.
The computational work was conducted using the resources of
the Northeast Parallel Architectures Center (NPAC) at Syracuse
University.

\figure{\label{fig:aval}
CDW velocity $v$ vs.\ time $t$ for a $128^{2}$ sample, with
$(E-E_{T})/E_{T}=0.98$ and $T=10^{-5}$ in dimensionless units.
Noise-triggered avalanches are visible above
the local thermal fluctuations.
The inset shows the avalanches which occur in the interval
between the two arrows; dark regions show where the phase
$\phii$ advances by more than $\pi$. In most of the volume
of the avalanches, the phase advances by an amount close to $2\pi$, while
in the light regions, the phase changes by much less than $1$.
}

\figure{\label{fig:smooth}
Plot of CDW velocity $v$ as a function of dimensionless temperature $T$
at the threshold field $\Et$, with smooth pinning potential.
The key indicates the sample size and number of samples averaged over.
The lines show slopes given by $\zeta/\tau$ for exponent values given in
the text for $d=2$ (dashed) and $d=3$ (dotted).
}
\figure{\label{fig:scalingfn}
Thermal rounding of the depinning transition for $d=2$ with
smooth potential: symbols show
scaled velocity $vT^{-\zeta/\tau}$ vs.\ scaled reduced field
$fT^{-1/\tau}$ for two different temperatures, using $\tau=1.5$ and
$\zeta=0.63$.
The solid line shows power law behavior $v\sim f^{\zeta}$, which
determines the scaling function $\bbar$ in the large $fT^{-1/\tau}$ limit.
}

\begin{references}
\bibitem{flrscf}
H.~Fukuyama and P.~A.~Lee, \PRB{17}, 535 (1977);
P.~A.~Lee and T.~M.~Rice, \PRB{19}, 3970 (1979);
K.~B.~Efetov and A.~I.~Larkin, Sov. Phys. JETP {\bf 45}, 1236 (1977);
L.~Sneddon, M.~C.~Cross and D.~S.~Fisher, \PRL{49}, 292 (1982).
\bibitem{review} For reviews, see
{\em Charge Density Waves in Solids},
edited by L.~P.~Gorkov and G.~Gr\"{u}ner (Elsevier, 1989);
G. Gr\"uner, Rev. Mod. Phys. {\bf 60}, 1129 (1988).
\bibitem{dsfone}
D.~S.~Fisher, \PRL{50}, 1486 (1983); \PRB{31}, 1396 (1985).
\bibitem{numerics}
For a review of numerical work, see
P.~B.~Littlewood, in \cite{review}.
\bibitem{thesis}
A.~A.~Middleton and D.~S.~Fisher, \PRL{66}, 92 (1991);
A.~A.~Middleton, Princeton University thesis, 1990 (unpublished);
D.~S.~Fisher and A.~A.~Middleton, to be published.
\bibitem{myersethna}
C.~Myers and J.~Sethna, to be published.
\bibitem{bhatt}
M.~O.~Robbins, J.~P.~Stokes, and S.~Bhattacharya, \PRL{55}, 2823 (1985);
S.~Bhattacharya, M.~J.~Higgins and J.~D.~Stokes,
\PRL{63}, 1503 (1989);
J.~McCarten, {\em et al}, \PRB{43}, 6800 (1991).
\bibitem{pslip}
M. Inui and S. Doniach, \PRB{35}, 6244 (1987); S.~H.~Strogatz,
C.~M.~Marcus, R.~M.~Westervelt, and R.~E.~Mirollo, \PRL{61},
2380 (1988).
\bibitem{sue}
S.~Coppersmith, \PRL{65}, 1044 (1990); S.~Coppersmith, to be
published.
\bibitem{nbnnote}
If a large part of the current is due to tearing of the CDW,
this should also be apparent in the relation between
the narrow-band-noise frequency and the CDW current, which is linear when
the CDW slides as a whole.
\bibitem{natterman}
P.~B.~Littlewood and R.~Rammal, \PRB{38}, 2675 (1988);
T.~Natterman, \PRL{64}, 2454 (1990);
J.~Toner, \PRL{64}, 2537 (1991).
\bibitem{pbl}
P.~B.~Littlewood, \PRB{33}, 6694 (1986);
P.~Sibani and P.~B.~Littlewood, \PRL{64}, 1305 (1990).
\bibitem{sncdsf}
S.~N.~Coppersmith and D.~S.~Fisher, \PRA{38}, 6338 (1988).
\bibitem{foot:units}
L.\ Pietronero and S.\ Strassler, \PRB{28}, 5863 (1983).
\bibitem{nanote}
For a mode $\sik$ which becomes unstable at a field $F_{k}$, the
eigenfrequency behaves as $\Lambda_{k}\sim(F-F_{k})^{\mu}$, with $\mu$
apparently having the trivial value $1/2$ \cite{thesis,sncdsf}, so that
$\chik$ does not diverge as $F\nearrow F_{k}$.
\end{references}
\end{document}